\documentclass[prb,preprint,showpacs]{revtex4}
\usepackage{epsf}
\usepackage{graphicx}
\begin{document}
\title{Quasiperiodic free electron metal layers} 
\author{A. K. Shukla,$^1$ R. S. Dhaka,$^1$ S. W. D'Souza,$^1$ Sanjay Singh,$^1$ D. Wu,$^2$\\ T. A. Lograsso,$^2$
 M. Kraj\v{c}\'{\i},$^{3,4}$ J. Hafner,$^4$ K. Horn,$^5$ and S. R. Barman$^1$}
\affiliation{$^1$UGC-DAE Consortium for Scientific Research, Khandwa Road, Indore, 452001, India.}
\affiliation{$^2$Ames Laboratory U.S. DOE, Iowa State University, Ames, Iowa 50011-3020, USA.}
\affiliation{$^3$Institute of Physics, Slovak Academy of Sciences, SK-84511 Bratislava, Slovak Republic}
\affiliation{$^4$Center for Computational Materials Science, Universit\"{a}t Wien,  A-1090 Wien, Austria}
\affiliation{$^5$Fritz-Haber-Institut der Max-Planck-Gesellschaft, D-14195 Berlin, Germany}

\begin{abstract}
Using electron diffraction, we show that free electron metals such as sodium and potassium form a highly regular quasiperiodic monolayer on the fivefold surface of icosahedral Al-Pd-Mn  and that the quasiperiodicity propagates up to the second layer in sodium. Our photoelectron spectroscopy results show that the quasicrystalline alkali metal adlayer does not exhibit a pseudogap near the Fermi level, thought to be  charactersitic for  the electronic structure of quasicrystalline materials. Calculations based on density functional theory provide a model structure for the quasicrystalline alkali metal monolayer and confirm the absence of a pseudogap. 
\end{abstract}



\maketitle 
\section{Introduction}
Quasicrystals are complex alloys of metallic elements exhibiting anomalous  properties such as high electrical resistivity, low thermal conductivity, low frictional coefficient and low surface energy.\cite{Shechtman84} In recent years, high quality quasicrystalline surfaces have been prepared and their electronic structure and morphology have been studied in detail. \cite{Gierer97,Rotenberg2000,Zheng04,Krajci05,Sharma2007} This has stimulated interest in the formation of quasiperiodic metallic films, using the quasicrystalline surface as template. Quasiperiodic films consisting of a single element have been achieved  only for a few cases (Sb, Bi, Cu, Sn) so far.\cite{Franke02,Ledieu04,Sharma05}  To date, there is no clear understanding for which  elements the interaction with the substrate might be sufficient to stabilize quasiperiodic order in the adlayer. Alkali metals like sodium or potassium are the best examples of nearly-free-electron metals and it is an intriguing question whether the growth of pseudomorphic films of these metals might be used to produce a two-dimensional (2D) quasiperiodic free-electron metal. 

Adsorption of alkali metals on metal surfaces has been a topic of active research because of interesting effects like  decrease of the work function, existence of collective multipole excitations and of quantum well states without a confining barrier.\cite{Bradshaw89,Barman01}  Alkali metal adsorption on the aluminum surface has been studied most extensively.\cite{Bradshaw89} The fivefold surface of icosahedral (i) Al-Pd-Mn has a similarity with the close-packed Al(111) surface, since it consists of a topmost dense Al-rich layer  separated by a distance of only 0.48~\AA~by a layer with  approximately 50$\%$ Al and 50$\%$ Pd, and the atomic density of these two layers together (0.136 atoms/\AA$^2$) is similar to that of Al(111).\cite{Gierer97} Our  x-ray photoelectron spectroscopy study of Na and K films on i-Al-Pd-Mn showed that alkali metal adlayers grow as a dispersed phase ({\it i.e.} not as condensed islands) up to one monolayer (ML).\cite{Shukla2006} Motivated by this work, alkali metal growth on i-Al-Pd-Mn has been studied theoretically using {\it ab initio} density functional methods and it was predicted that Na and K monolayers  exhibit quasiperiodic order.\cite{Krajci07}

An interesting feature of the electronic structure of many icosahedral quasicrystals is the formation of a \verb+"+pseudogap\verb+"+, {\it i.e.} of a minimum in the electronic density of states (DOS) close to the Fermi level ($E_F$). Many of the outstanding properties such as their anomalous transport properties have been attributed to the existence of the pseudogap.\cite{Stadnik99} Photoelectron spectra show a rounding of the Fermi edge, which has been interpreted as the signature of such a DOS minimum. \cite{Stadnik99,Wu95,Stadnik96,Neuhold98} It is an interesting question whether the strong coupling between adatoms and substrate necessary to impose quasiperiodic order also leads to the formation of a pseudogap in the DOS of the 2D adlayer. In this work, we show that  Na and K quasiperiodic films can be prepared on the fivefold surface of i-Al-Pd-Mn, and report on the characterization of their geometric and electronic structures using low-energy electron diffraction (LEED) and photoemission spectroscopy (PES). We find good agreement between the experimental results and the theoretical predictions.\cite{Krajci07}

\section{Experimental}
Single grain i-Al-Pd-Mn quasicrystal with bulk composition Al$_{69.4}$Pd$_{20.8}$Mn$_{9.8}$ was grown using the Bridgman technique.  The polished i-Al-Pd-Mn  surface was cleaned by repeated cycles of Ar$^+$  sputtering (1.5~keV, 45-60~minutes) and annealing (2-2.5 hours) up to 870-900~K to produce large flat atomic surface. LEED  and PES experiments have been performed using 4-grid rear view ErLEED optics, and a Phoibos 100 electron energy analyzer, respectively. The base pressure of the experimental chamber was 6$\times$10$^{-11}$ mbar. He~I radiation (h$\nu$=\,21.2 eV) was used for the photoemission experiments. The resolution of the PES is about 0.1~eV at the  measurement temperature (130~K). The composition of the quasicrystalline surface is Al$_{74}$Pd$_{21}$Mn$_{5}$, as calculated from area under the core-level peaks (Al 2$p$, Pd 3$d$, Mn 2$p$), {\it i.e.}  very close to the bulk composition. Na and K were deposited from well-degassed commercial SAES getter sources. The substrate is held at 130~K to avoid possible surface alloying that is observed for alkali metals at room temperature.\cite{Barman2001} During deposition, the chamber pressure rose to 9$\times$10$^{-11}$ mbar. The thickness of the adlayers has been calculated from the area under the (Na 1$s$ and K 2$p$) and (Al 2$p$) core-level peaks, as in our previous work.\cite{Shukla2006,Shukla04} The atomic density of a completed monolayer is 0.067 atoms/\AA$^2$, corresponding to a coverage of $\Theta=0.50$, i.e roughly equal to the coverage of the $(4\times 4)$ phase of Na on Al(111).\cite{Bradshaw89}

\section{Results and Discussion}
The structure of the alkali metal overlayer on i-Al-Pd-Mn as seen in LEED, recorded with 78~eV electron energy, is shown in Fig.~1. At such energies, LEED is sensitive primarily to the topmost layers due to the small mean free path of the electrons.\cite{Tanuma91} In the LEED pattern of the clean i-Al-Pd-Mn surface (Fig.~1(a)), an inner ring consisting of 10 sharp spots is visible. A closer inspection of the spot intensities reveals two inequivalent sets of five spots, which agrees with the expected 5-fold symmetry. An outer ring with similar arrangement of the inner ring is also visible. The LEED pattern is in agreement with previous work on the five fold i-Al-Pd-Mn surface.\cite{Gierer97,Franke02} 

\begin{figure}[htbp]
\includegraphics[width=13cm]{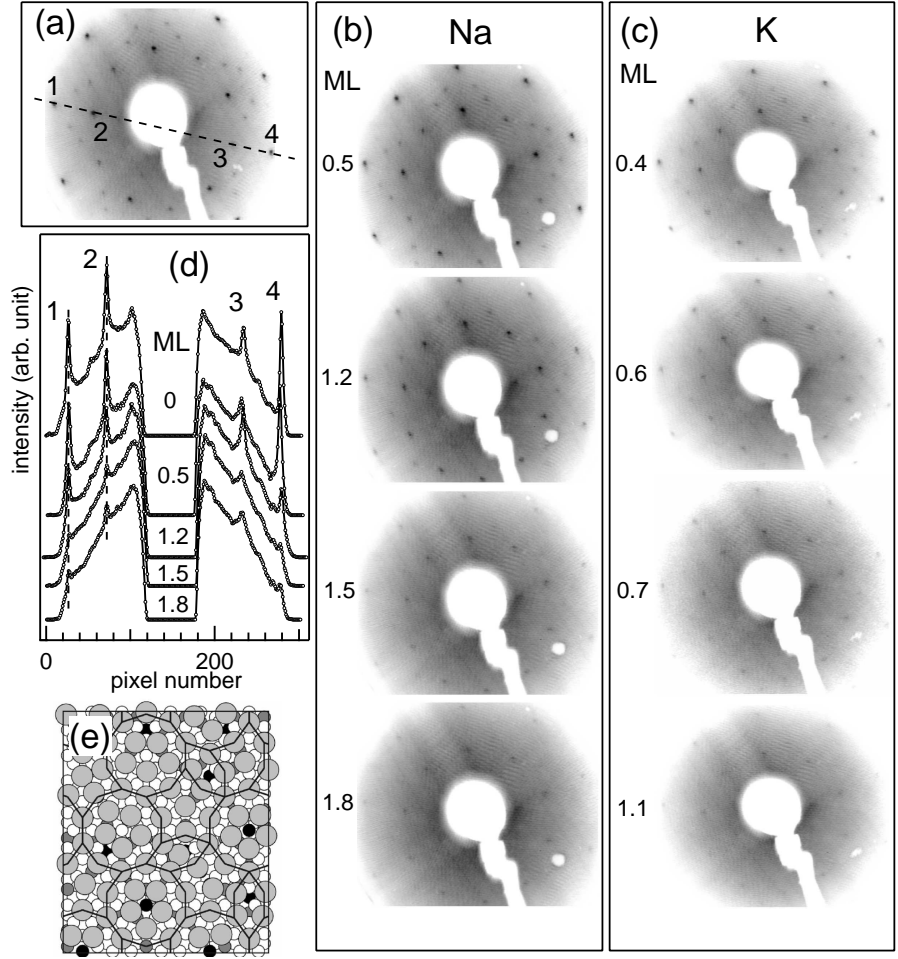}
\caption{(a) LEED pattern of a clean i-Al-Pd-Mn surface. Diffraction spots are numbered to facilitate their identification in the intensity profiles (shown in (d)) along the dashed line going through different spots. (b) and (c) show LEED patterns measured at 130~K for different Na and K coverages on i-Al-Pd-Mn, respectively. Images are shown in inverted gray scale where black indicates highest brightness. All the LEED patterns have been recorded at 78 eV electron beam energy. (d) Intensity profile as a function Na coverage along the dashed line shown in (a). (e) The atomic structure of an adsorbed K monolayer on the Al-Pd-Mn substrate where Al: small open circles, Pd: small gray circles, Mn: small black circles, and K: large gray circles. The quasiperiodic ordering is represented by the DHBS tiling, cf. text.}
\end{figure}

The fivefold symmetric LEED pattern persists upon dosing with Na or K, as shown in Figs.~1(b) and 1(c). The positions and widths of the LEED spots or the background remain mostly unchanged up to 1.2~ML Na coverage (Fig.~1(b)).  This clearly shows that the Na adlayer grows pseudomorphically and  adopts the quasicrystalline symmetry of the substrate. For 1.5~ML of Na, the spot intensities slightly decrease and the background intensity increases. For 1.8~ML of Na, the spot intensities decrease further, with a concomitant increase in background, and the LEED spots intermediate between the inner and outer rings are hardly visible. This indicates that the quasiperiodic order in the Na adlayer decreases above 1.2~ML. However, a distinctive 5-fold LEED pattern exists up to 1.8~ML Na with no change in LEED spot positions. This suggests that pseudomorphic growth continues even for higher Na coverages ($>$1.2~ML). The results for K layers are slightly different as shown in Fig.~1(c): the LEED spots for different coverages of K  larger than 0.6~ML decrease in intensity and the intermediate spots disappear. However, a 5-fold LEED pattern is visible up to 1.1~ML. Thus, K/i-Al-Pd-Mn films grows as a well ordered quasicrystalline film up to about 0.6~ML, while for higher coverages the degree of quasicrystalline order diminishes. 

The intensity profiles of the diffraction spots are often used to obtain information about the degree of surface order.\cite{Henzler82} Splitting or significant broadening of the LEED spots indicates the presence of surface defects, especially steps of different heights.\cite{Van79} Fig.~1(d) shows the intensity profile of the LEED spots as a function of Na coverage along a dashed line joining two of the most intense spots: 1 and 4 in Fig.~1(a). The peaks in intensity profile correspond to the observed LEED spots, which are numbered in Fig.~1(a). Flat portions in the profiles are due to the shadow of the electron gun.

Important observations can be made from the intensity profiles. First, with increasing coverage of Na, there is no change in the LEED spot positions. The ratio of the distances between the LEED spots forming the inner and outer ring from the center (specular spot) remains equal to  the Golden Mean, $\tau \approx$1.61). Second, a splitting of the LEED spots is not observed indicating that the adlayer is atomically smooth.  Finally, we do not observe the emergence of any extra spots after Na adsorption up to the highest coverage. Similar intensity profiles are obtained for K adlayers.

It is well known that presence of step arrays on the surface results in the splitting of LEED spots.\cite{Ellis68,Henzler70,Janzen99} Occurence of the periodically alternating single and splitted integer order spots with increasing electron beam energies ({\it i.e.} increasing radius of Ewald's sphere) has been attributed to an out-of-phase scattering of subsequent terraces and a formula was derived to determine the characteristic electron energies for sharp LEED spots.\cite{Henzler70,Janzen99} Figure~2 shows the LEED patterns of 1.2~ML Na recorded at various electron beam energies. LEED spots of 5-fold diffraction pattern continuously converge towards specular spot (hidden by the shadow of the electron gun) with increasing incident beam energies and we do not observe the splitting of the LEED spots at any electron energy. It clearly excludes the presence of step defects in the adlayer.

\begin{figure}[htbp]
\includegraphics[width=13cm]{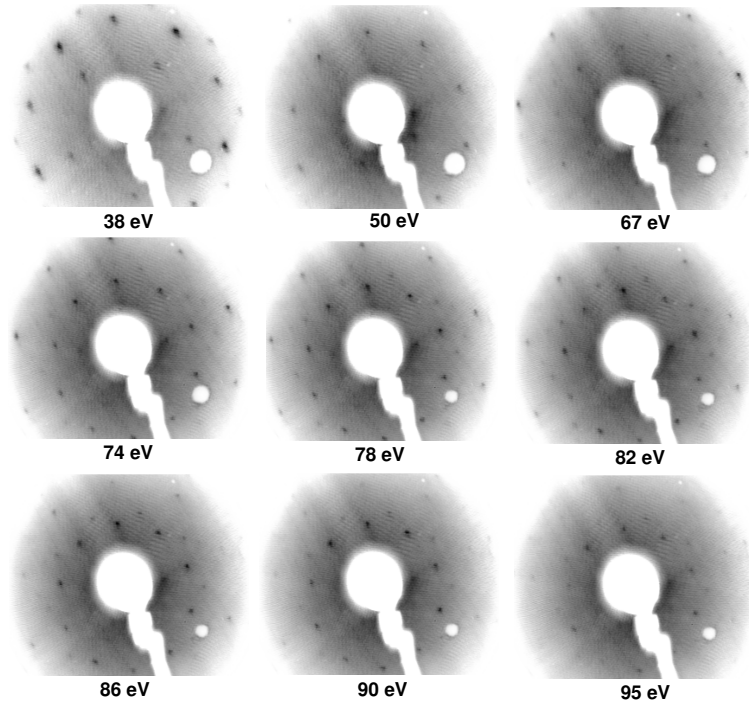}
\caption{LEED patterns of 1.2~ML Na grown on i-Al-Pd-Mn surface at 130~K at several incident electron beam energies. Images are shown in inverted gray scale where black indicates highest brightness. All the LEED patterns have been recorded at 130~K.}
\end{figure}

If an adlayer has different domains or twinned structures with a specific orientational relationship with the substrate, then it results in the appearance of multiple spots in the LEED pattern. There are reports of rotational epitaxy for different metals (Al, Ag, Fe, Ni) grown on i-Al-Pd-Mn, where the adlayer forms as five crystalline domains reflecting the substrate symmetry and the formation of such domains results in the extra spots in the LEED pattern of metal films on the quasicrystalline surfaces.\cite{Sharma2007,Bollinger01} The LEED patterns for 1.2~ML Na recorded at various electron energies (Fig.~2) do not show the presence of such multiple spots. We also do not observe the multiple spots in LEED pattern of K adlayers. This eliminates the possibility of the formation of twinned nano-crystallites or crystalline domains in the alkali metal adlayers. 

The geometric structure of a five fold surface of i-Al-Pd-Mn is well described by a P1 tiling consisting of pentagons, pentagonal stars, boats and thick and thin golden rhombi,
\cite{Krajci05,Krajci07,Papado02} as confirmed also by atomically resolved scanning tunneling microscopy.\cite{Krajci06} The surface consists of two closely spaced atomic layers with a vertical separation of only 0.48~\AA. The top layer is occupied by a majority of Al atoms and $\approx$4 at.\% of Mn, while the second layer is composed in a ratio of about 1:1 by Pd and Al atoms with a small concentration of  Mn atoms. Half of the vertices of the P1 tiling are decorated by these Pd atoms, while most of the remaining vertices correspond to Pd atoms in deeper layers. In the surface plane these sites are surrounded by complete or incomplete pentagons of Al atoms. 

Our mapping of the potential energy surface of an alkali atom on the i-Al-Pd-Mn surfaces using ab-initio DFT techniques (we used the VASP code\cite{vasp}) demonstrates that the vertices of the P1 tiling (decorated by Pd atoms located 0.48~\AA~below the surface and characterized by minima in the surface electron-density) are indeed the energetically most favorable adsorption sites. Hence these sites were chosen to form the skeleton of the adlayer structure, the interior of the tiles being decorated in a way compatible with the overall fivefold symmetry and an atomic density of 0.66 atoms/\AA$^2$ (corresponding to $\Theta$\,=0.5). Relaxation of the starting structure under the action of the ab-initio calculated forces resulted in a spontaneous rearrangement of the adatoms in a highly regular quasiperiodic structure different from the initial structure, with atomic positions conforming with a decagon-hexagon-boat-star (DHBS) tiling. The alkali atoms initially placed onto the vertices of the P1 tiling remain in their positions, but the atoms decorating the interior of the tiles move in such a way as to form an even more regular structure - the relaxed equilibrium structure in a K-adlayer is shown in Fig.~1(e). The shortest ideal distance between atoms forming the pentagonal motifs in the relaxed adlayer is 3.60~\AA, only in the boat tile a few distances of 3.41\,\AA~ are observed. Hence this structure is ideally suited for a regular adlayer of Na atoms (atomic diameter 3.7\,\AA~ as calculated for bulk Na), while a K adlayer (atomic diameter 4.6\,\AA)~ is already slightly over-packed, as reflected by a somewhat larger corrugation of the adlayer. Thicker films may be formed by depositing a second layer of adatoms into hollow sites of the first monolayer. A stable quasiperiodic bilayer structure is formed if the coverage in the second layer is reduced to $\Theta$=\,0.39, while quasiperiodic order decreases rapidly on addition of further atomic layers. 

The theoretical predictions based on density functional theory are consistent with
our experimental observation of quasiperiodic ordering in a Na overlayer at T=\,130~K, which gradually disappears above 1.8 ML. The calculations show that both K and Na monolayers exhibit the same quasiperiodic pattern, however, the lateral ordering in a K monolayer is less regular and the monolayer is more corrugated. In the experiment, it is seen that the quasiperiodicity of a K film is affected even below monolayer thickness, which is not the case for Na. The different degree of ordering in Na and K layers arises from the different atomic size: In the idealized DHBS structure the distances between the K atoms are already somewhat compressed, leading to a stronger corrugation and a less perfect quasiperiodic order and these effects become even more pronounced  with increasing coverage.

\begin{figure}[htbp]
\centering
\includegraphics[width=13cm]{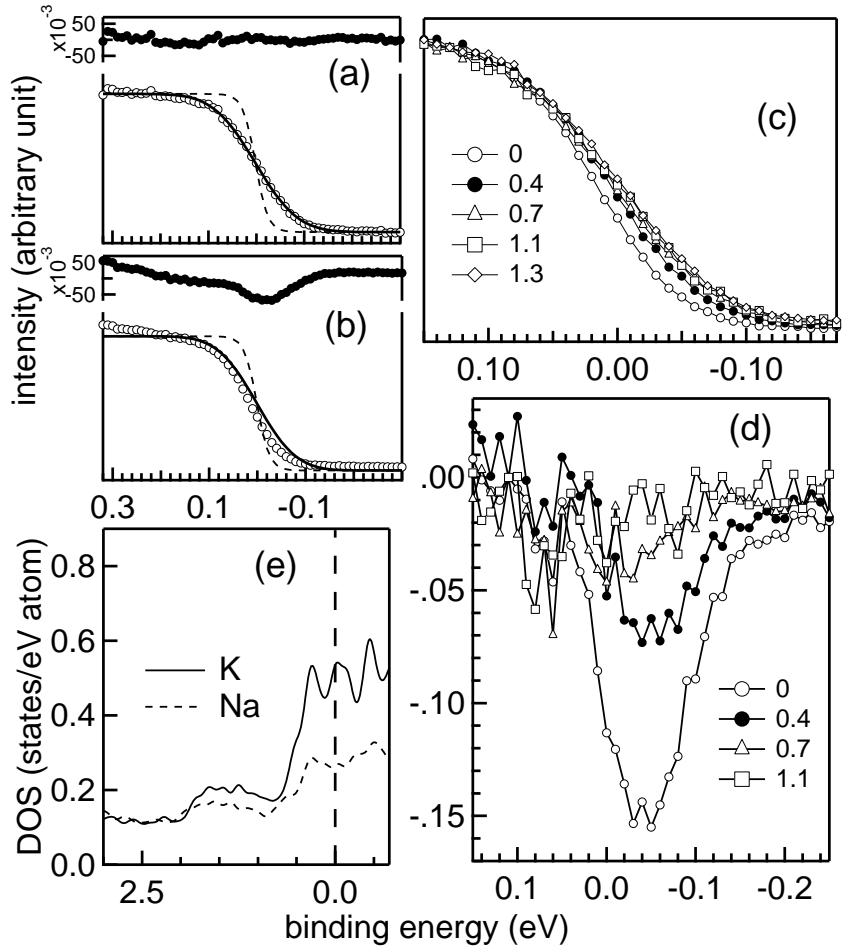}
\vskip -0.5cm
\caption{Photoemission spectra in the near $E_F$ region (open circles) of (a) 1.3~ML K coverage on i-Al-Pd-Mn, and (b) clean i-Al-Pd-Mn surface at 130~K (open circles). The fitted curve (thick solid line) and Fermi function (dashed line) are also shown. Residual of fitting is shown at the top of each spectra. (c) Closeup of near Fermi edge region as function of K coverage. Spectra are normalized to the same height at 0.1~eV. (d) The difference spectra are obtained by subtracting the near $E_F$ region spectrum of 1.3~ML K from those of other K coverages shown in (c). (e)  The calculated DOS of Na and K monolayer adsorbed on i-Al-Pd-Mn.} 
\end{figure}

Having demonstrated the formation of a 2D quasiperiodic system consisting of a single free electron metal, we are in a position to compare its electronic structure to that of ordinary 3D bulk quasicrystals.
~Here, the central question is whether the pseudogap structure of the electronic DOS of the substrate is also imprinted upon the 2D adlayer. To resolve this question, we have performed electronic structure calculations and PES experiments covering the region around $E_F$ (Fig.~3). If a pseudogap at the Fermi level exists, photoemission spectrum near $E_F$ cannot be modeled simply by a Fermi function;\cite{Wu95,Stadnik99,Stadnik96,Neuhold98} rather the measured intensities show a slow  decrease on approaching $E_F$. By comparing the spectral function near $E_F$ for clean i-Al-Pd-Mn and progressively higher coverages of K, we show that the signature of a pseudogap, clearly found in i-Al-Pd-Mn is absent in the quasicrystalline K layer. We analyze the spectra for a 1.3~ML K film and a clean i-Al-Pd-Mn surface by fitting with a Fermi function convoluted with the instrumental Gaussian broadening (Figs. 3(a,\,b)). Least square error minimization is performed by freely varying the position, intensity of the Fermi function and the full width at half maximum of the Gaussian. 
~The residual at the top of Fig.~3(a) shows the good quality of the fit for 1.3~ML of K, indicating that no DOS minimum at the Fermi level exists for the K adlayer. On the other hand, because of a pseudogap,  the residual exhibits a dip in clean i-Al-Pd-Mn (Fig.~3(b)). This dip, which is the signature of the pseudogap, appears slightly above $E_F$, in agreement with earlier literature.\cite{Neuhold98,Krajci05a}

The absence of a pseudogap in alkali-metal adlayers on i-Al-Pd-Mn is clearly observed from the coverage-dependent spectra (Fig.~3(c)). The spectral intensity slightly above $E_F$ gradually increases with increasing coverage of K. The difference spectra in Fig.~3(d), obtained by subtracting the 1.3~ML K/i-Al-Pd-Mn spectrum from those with lower K coverage, 
~unambiguously show the progressive filling up of the pseudogap with increasing K coverage. Note that at  a coverage of 0.7~ML K, where the LEED data in Fig.~1(c) show that the layer is clearly quasiperiodic, the PES shows that no pseudogap exists. These experimental observations are in good agreement with DFT predictions. Fig. 3(e) shows the calculated DOS for Na and K monolayers on i-Al-Pd-Mn around $E_F$.  For a Na monolayer we find a continously increasing DOS, while for a K monolayer we even find a strongly increasing DOS at binding energies below 0.5 eV. This is the result of compressed K\,-\,K distances leading to an incipient occupation of K 3$d$ states (as also in compressed bulk K). 

This is an interesting finding because the existence of a pseudogap near $E_F$ is sometimes considered as an important factor contributing to the stability of the quasicrystalline phases (although it is evident that pseudogaps exist not only in quasicrystals, but also in many crystalline alloys such as Al-Mn and even amorphous alloys.\cite{Krajci93,Shukla08}) In any case, the existence of a deep pseudogap is an indication of a considerable degree of covalency in the interaction between the Al and the transition-metal atoms.\cite{Krajci03} Previous investigations (both theoretical\cite{Krajci05a} and experimental\cite{Fournee00}) of the surface electronic structure of i-Al-Pd-Mn have demonstrated that the metallic character of the bonding is enhanced at the surface and that the pseudogap is partially leveled out as a consequence of a strong relaxation of the near-surface Al atoms. The present work extends these studies to the electronic structure of a well-ordered quasiperiodic adlayer and we find that the pseudogap has disappeared. This has important consequences and the mechanism stabilizing the quasiperiodic structure of the films\,-\,quasiperiodicity is imprinted on the first monolayer by the strong binding of the adatoms in surface charge-density minima. Atoms in the second monolayer are also adsorbed in hollow sites of the first layer, but because the binding between the alkali atoms is much weaker than with the substrate, the quasiperiodic order is gradually lost in multilayers. 

\section{Conclusion}
We have performed low energy electron diffraction, ultra-violet photoemission and {\it ab-initio} density functional theory calculations to show that free electron metals such as sodium and potassium form a highly regular quasicrystalline monolayer on i-Al-Pd-Mn. In case of Na, quasiperiodicity propagates up to the second layer.  The existence of quasicrystalline alkali metal adlayers and that it propagates to the second layer in Na is in good agreement with the previously published density functional theory results. Photoelectron spectroscopy results show that the pseudogap of the i-Al-Pd-Mn substrate is gradually filled up as alkali metal is deposited. Quasicrystalline alkali metal adlayer does not exhibit a pseudogap near the Fermi level, thought to be  charactersitic for  the electronic structure of quasicrystalline materials. The density of states of an quasicrystalline alkali metal monolayer on i-Al-Pd-Mn calculated by density functional theory  also shows the absence of a pseudogap.

\section{Acknowledgment}
Fundings from  Max Planck Institute, Germany, Department of Science and Technology, India, Ramanna Research Grant and  US Department of Energy Basic Energy Sciences  are gratefully acknowledged.

\end{document}